\documentclass[sigconf,natbib=false]{acmart}
\usepackage{graphicx} 
\usepackage{array}
\usepackage{multirow}
\PassOptionsToPackage{table}{xcolor}
\usepackage{enumitem}
\usepackage{xspace}
\usepackage[size=\tiny]{todonotes}
\usepackage[final,commandnameprefix=ifneeded]{changes}
\newif\ifshowtodos
\showtodosfalse 

\ifshowtodos
  \providecommand\redtodo[1]{\textcolor{red}{#1}}
  \providecommand\bluetodo[1]{\textcolor{blue}{#1}}
  \providecommand\hardc[1]{\textcolor{purple}{#1}}
\else
  \providecommand\redtodo[1]{}
  \providecommand\bluetodo[1]{}
  \providecommand\hardc[1]{#1}
\fi

\newcommand{\CodexMedianTimeToMergeMinutes}{0.5\xspace}

\newcommand{\HumanMedianTimeToMergeHours}{0.4\xspace}

\newcommand{\HumanPctMergedUnderTenMin}{43.3\%\xspace}

\RequirePackage[
  datamodel=acmdatamodel,
  style=acmnumeric,
  citestyle=numeric-comp,
  natbib=true,
  sorting=none
  ]{biblatex}
\title{Investigating Autonomous Agent Contributions in the Wild: Activity Patterns and Code Change over Time}

\author{Razvan Mihai Popescu}
\email{r.m.popescu@tudelft.nl}
\affiliation{%
  \institution{Delft University of Technology}
  \city{Delft}
  \country{The Netherlands}
}
\author{David Gros}
\email{dgros@ucdavis.edu}
\affiliation{%
  \institution{University of California, Davis}
  \city{Davis}
  \country{USA}
}

\author{Andrei Botocan}
\email{a.botocan@student.tudelft.nl}
\affiliation{%
  \institution{Delft University of Technology}
  \city{Delft}
  \country{The Netherlands}
}

\author{Rahul Pandita}
\email{rahulpandita@github.com}
\affiliation{%
  \institution{GitHub}
  \city{Denver}
  \country{USA}
}

\author{Prem Devanbu}
\email{ptdevanbu@ucdavis.edu}
\affiliation{%
  \institution{University of California, Davis}
  \city{Davis}
  \country{USA}
}

\author{Maliheh Izadi}
\email{m.izadi@tudelft.nl}
\affiliation{%
  \institution{Delft University of Technology}
  \city{Delft}
  \country{The Netherlands}
}

\begin{document}



\begin{abstract}
 
The rise of large language models for code has reshaped software development. Autonomous coding agents, able to create branches, open pull requests, and perform code reviews, now actively contribute to real-world projects. Their growing role offers a unique and timely opportunity to investigate AI-driven contributions and their effects on code quality, team dynamics, and software maintainability.
In this work, we construct a novel dataset of approximately $110,000$ open-source pull requests, including associated commits, comments, reviews, issues, and file changes, collectively representing millions of lines of source code. We compare five popular coding agents, including OpenAI Codex, Claude Code, GitHub Copilot, Google Jules, and Devin, examining how their usage differs in various development aspects such as merge frequency, edited file types, and developer interaction signals, including comments and reviews. Furthermore, we emphasize that code authoring and review are only a small part of the larger software engineering process, as the resulting code must also be maintained and updated over time. 
Hence, we offer several longitudinal estimates of survival and churn rates for agent-generated versus human-authored code. 
Ultimately, our findings indicate an increasing agent activity in open-source projects, although their contributions are associated with more churn over time compared to human-authored code.

\end{abstract}
\maketitle

\section{Introduction}

In $2021$, early evidence emerged showing the ability of large pre-trained language models (LLMs) to write complete units of code. This was illustrated through the HumanEval dataset~\cite{chen2021he}, which evaluates the task of generating relatively simple Python programs~\footnote{The median is only seven lines-long reference solutions.}.
They reported that GPT-3 could not solve any of the problems, while their proposed Codex model was able to solve $28.8$\% of the problems. Other contemporaneous work confirmed that LLMs were capable of generating methods in various programming languages~\cite{austin2021programsynthesislargelanguage, hendrycks2021measuringcodingchallengecompetence, multiple}.

The $2022$ release of systems such as GitHub Copilot for code completion and ChatGPT for general dialog accelerated the broader use of LLMs for software. However, the assistive nature and limited environmental integration of these early systems have made it challenging to observe and assess their use in the wild, with most research focusing on acceptance rates, which, sadly, fail to capture representative software quality attributes~\cite{code4me, intellicodecompose, productivitycopilot}.
Since then, AI assistants have become increasingly embedded in developer workflows, contributing to higher levels of user satisfaction and productivity~\cite{productivity1, productivity2, productivity3, productivity4, productivity5, jetbrains2024}. 
Recent works report that over $90$\% of coding assistants have been released in the past two years~\cite{productivity3}, while $84$\% of developers are already using or plan to incorporate AI tools in their development process~\cite{stackoverflow2025}. 

With this wave of progress, the software engineering field is transitioning into an AI-native phase, where development is guided by human intent and collaborations with autonomous AI-coding partners~\cite{hassan2024ainativesoftwareengineeringse, li2025riseaiteammatessoftware}.
Such emerging agents move beyond passive AI-assistants, being characterized by three core features, including autonomy, expanded scope, and engineering practicality. With improved tooling, these agents can now actively manage and execute development workflows from requirements to implementation~\cite{dong2025surveycodegenerationllmbased}.

The fast-paced and sheer volume of contributions produced by agentic systems, which already accounts for $10$\% of public Pull Requests (PR) on GitHub\added{~\cite{logicstar2025dashboard}}, creates the need to examine their activity and understand how these systems might influence factors such as code maintainability, which is traditionally the most resource-intensive phase of the software development life-cycle~\cite{10.1145/3611643.3616339}. Unlike traditional assistants or dialog systems, where mixed developer and AI activity made provenance difficult to trace without access to specialized telemetry~\added{or self-admitted usage artifacts~\cite{tufano}}, agent system's enhanced environment integration enables relatively~\added{easier identification of code generated with heavy AI assistance, using their tell-tale signals~\cite{logicstar2025dashboard, li2025riseaiteammatessoftware, watanabe2025useagenticcodingempirical}}.

In this work, we take advantage of this distinction to mine GitHub PRs written by autonomous coding agents. Previous studies in this area have examined the role of models such as ChatGPT in software development~\cite{tufano, watanabechat, sun2025doesaicodereview}, while others have explored the impact of reactive AI assistants on code maintainability through controlled user studies~\cite{borg2025echoesaiinvestigatingdownstream} or synthetic benchmarks~\cite{yetiştiren2023evaluatingcodequalityaiassisted}. However, such approaches remain limited in capturing how these models interact with existing code due to the isolated nature of their evaluation protocols.
More recent work investigating the integration of agentic coding in collaborative development remains centered on high-level contribution dynamics, such as differences in PR purpose, acceptance rates, revision effort, or review behavior~\cite{logicstar2025dashboard, li2025riseaiteammatessoftware, watanabe2025useagenticcodingempirical}. More importantly, most of these investigations focus on a \textit{limited} subset of PRs and are often restricted to \textit{highly popular repositories}~\cite{li2025riseaiteammatessoftware} or a single coding agent~\cite{watanabe2025useagenticcodingempirical}. 
We highlight how, unlike human PRs, most agent PR work is concentrated in very low-star~\added{repositories}, making a broad view important for understanding these systems.
Our study complements current literature by providing a \textit{broader} and \textit{deeper} view of the agentic landscape.

Additionally, we contribute new directions to this understanding by studying the characteristics of agentic activity within collaborative development, with a particular focus on change signals associated with the quality of agent-authored code. This includes a first-of-its-kind longitudinal study of the evolution of agent-authored PR code, focusing on five coding agents, including OpenAI Codex, GitHub Copilot, Claude Code, Google Jules, and Devin. Specifically, our main research questions (RQs) are: 
\begin{itemize}[label={RQ}]
    \item[\textbf{RQ1:}] What is the difference between agent-authored and human-authored PRs in shaping collaboration and development progress? 
    \item[\textbf{RQ2:}] How do agent-authored PRs influence the trajectory of code maintenance over time compared to human-authored ones?
\end{itemize}

Our findings indicate that although agent activity is growing in open-source projects, particularly in lower-starred repositories, their contributions are correlated with more extensive code changes over time compared to human-authored code.

Our contributions are as follow: 
\begin{itemize}
    \item We curate a large-scale dataset comprising over $110,000$ PRs and related metadata from public GitHub data, capturing the activity and collaboration of humans and coding agents in open-source software development spanning several months. 
    \item We examine the difference in activity dynamics between agent- and human-authored PRs across various dimensions, including merge rates, merge latency, complexity of changes, file composition, commit and review density, and repository characteristics. 
    \item We assess the subsequent evolution and stability of the code contributed by agents and humans using indicators such as \textit{survival rate} and \textit{code churn}.
    \item We release our dataset to support further research on agentic software development~\footnote{https://huggingface.co/datasets/AISE-TUDelft/MOSAIC-agentic-3m}.
\end{itemize}

\section{Related Work}
Below, we review prior work on AI assistants’ evolution into autonomous agents, their evaluation, and their real-world activity and collaborative interactions with human developers. 

\subsection{From Assistants to Agents}

AI assistants are~\added{typically} defined by their assistive and predictive capabilities, integrated within a developer's workflow,~\added{for example,} as IDE extensions~\added{\cite{li2025riseaiteammatessoftware}}. 
These systems can operate as pair programmers designed to support developers across various coding tasks, such as generating code~\added{\cite{code4me}}, or identifying existing issues or vulnerabilities~\added{\cite{aibughunter}}. 
Early coding assistants including Code4Me~\cite{code4me} or IntelliCode Compose~\cite{intellicodecompose}, powered by generative transformer models such as GPT-C and InCoder, were able to perform short-sequence token predictions and automatically complete entire lines of syntactically correct code in several programming languages. 
Similarly, GitHub Copilot and Amazon CodeWhisperer could transform natural language comments into lines or entire blocks of source code taking into account code-context~\cite{yetiştiren2023evaluatingcodequalityaiassisted}, while AIBugHunter was able to locate and classify vulnerability types, estimate their severity, and even suggest potential vulnerability fixes to developers~\cite{aibughunter}. 
However, all these AI assistants require user confirmation and oversight at each step in their interaction model. Additionally,  despite evidence of increased developer productivity and satisfaction~\cite{productivitycopilot}, in many cases these coding assistants still lack a clear understanding of contextual clues, leading to the introduction of code smells that can cause technical debt in the long term~\cite{productivity3, yetiştiren2023evaluatingcodequalityaiassisted}.

The improving ability of LLMs, both on code-generation and instruction-following, has fueled a transition from reactive coding assistants to more proactive agents that can reason over broader project contexts. 
These agents are able to autonomously perform multi-step task planning, take actions, run self-evaluations, and coordinate across multiple tools~\cite{survey1}. A growing number of agentic systems now demonstrate this kind of autonomous coordination across the software development cycle~\added{\cite{survey1}}. RepoAgent has been used to proactively generate and maintain comprehensive documentation for entire code repositories~\cite{repoagent}, while RepairAgent, which leverages a finite state machine to mimic human debugging states, has shown promising results in fixing software bugs~\cite{repairagent}. Furthermore, SWE-agent~\cite{swe-agent} and OpenHands~\cite{openhands} were designed to equip LLMs with specialized interfaces to transform them from reactive predictors into autonomous agents capable of complex, multi-step actions.~\added{Accordingly,} these models can navigate entire repositories, execute programs, or edit code files~\added{\cite{swe-agent, openhands}}; in addition, OpenHands also enables the delegation of tasks to other specialized agents to solve challenging problems. 

Additionally, there is a rise of multi-agent systems as well, which mimic the task distribution, specialized roles, and~\added{coordination} activity of real developer teams. AgentCoder~\cite{agentcoder} allows agents to take on a different role, including programmers, test designers, and test executors. Similarly, for MaintainCoder~\cite{maintaincoder}, which focuses on ensuring high code maintainability, agents are responsible for tasks ranging from requirement analysis and design pattern selection to framework design, evaluation, and code optimization. Lastly, CodeSim~\cite{codesim} addresses program synthesis with its agents focusing on planning, coding, and debugging, introducing a unique simulation-driven approach for evaluating agents' work internally, mirroring how humans visualize and refine algorithms.    

These agents not only add a layer of autonomy and persistence over their forerunners, but also possess the ability to adjust their behavior over the observed effects of their own actions while maintaining progress towards the user-provided goal. This is possible through feedback looping, either by internal self-correction, collaboration with humans, or negotiation within multi-agent systems. Nevertheless, this shift from heavy developer input to increasing model autonomy can encourage superficial reliance, potentially leading over time to buggy, unmaintainable, or security-vulnerable code. 

\subsection{Agent Evaluations}
To assess AI assistants based on output quality, usability, user satisfaction, or productivity, most prior studies rely on controlled user experiments that evaluate the relative performance of these systems in regards to the developer experience and task goals~\cite{barke, becker2025measuringimpactearly2025ai}. Although this evaluation setting is strongly related to the operational nature of these tools, it suffers from low external validity, not entirely capturing the complexity, pace, and~\added{nuances} of real-world development. 

When it comes to agentic systems, static benchmarking represents the primary method to evaluate their capabilities in isolation. Pioneering benchmarks, including HumanEval, MBPP, APPS, and Defects4J~\cite{defects4j}, are still in the community's attention, despite known issues related to data contamination, erroneous ground truths, or saturation~\cite{contamination, ionstoica}. More recently, a plethora of agentic benchmarks have emerged, targeting different software engineering tasks. SWE-Bench evaluates the ability of agents to solve entire GitHub issues by interacting with repositories~\cite{swe-agent}. Similarly, SWT-Bench, built on top of SWE-Bench, is specifically designed to evaluate the ability to generate bug reproduction tests~\cite{swtbench}. Other benchmarks, such as AgentBench, assess agents’ decision-making capabilities in diverse environments, including bash scripting, interacting with real SQL databases, or web browsing~\cite{agentbench}. Meanwhile, MaintainBench measures agents’ capability to produce maintainable code through requirement evolution cycles~\cite{maintaincoder}. Although these benchmarks provide controlled and reproducible measures of agent performance, they capture only narrow aspects of their development abilities within idealized conditions. This further highlights the need for evaluations that can capture authentic developer-agent interaction and operational project dynamics.  

\begin{table}[t]
\caption{Search signals used to filter~\added{PRs per} coding agent.}
\label{tab:search_signals}
\Description{Search signals used to filter PRs per coding agent.}
\centering
\begin{tabular}{p{0.13\textwidth} >{\centering\arraybackslash}p{0.30\textwidth}}
\toprule
\textbf{Coding Agent} & \textbf{Search Signal} \\ \midrule
\textbf{OpenAI Codex} & head:codex/ \\
\textbf{GitHub Copilot} & head:copilot/ \\
\textbf{Google Jules} & author:google-labs-jules[bot] \\
\textbf{Devin} & author:devin-ai-integration[bot] \\
\textbf{Claude Code} & (``Co-Authored-By: Claude'' OR ``Generated with Claude Code'') \\
\bottomrule
\end{tabular}
\end{table}

\subsection{Agents In The Wild}

The growing integration of LLMs into development workflows has motivated efforts to examine their real-world footprint, particularly how these models contribute to and shape activity within repositories. Previous investigations of pull requests, commits, and issues show that developers employ ChatGPT to automate various tasks, including documentation, bug fixing, and code generation~\cite{tufano}. Other studies have examined how LLMs and AI-based tools participate in code reviews, either by assisting developers in providing feedback and alternative solutions~\cite{watanabechat}, or by automatically generating review comments within GitHub workflows~\cite{sun2025doesaicodereview}. These findings highlight both the growing adoption and the variable effectiveness of AI-generated reviews, emphasizing the need for careful tool design and a better understanding of their broader impact.

Following this, other studies have focused on the effects of integrating coding agents into software development workflows. Using the AIDev-Pop dataset, containing $7,122$ PRs from GitHub repositories with more than $500$ stars and involving agents such as OpenAI Codex, Devin, GitHub Copilot, Cursor, and Claude Code, this work explores productivity patterns~\cite{li2025riseaiteammatessoftware}. They analyze factors such as resolution and turnaround times, acceptance rates, and structural properties of agent-generated code, including complexity and the nature of the changes. Their results showed that agents can drastically accelerate developer output (e.g., one developer submitted $164$~\added{a}gentic-PRs in three days versus $176$~\added{h}uman-PRs over three years), and PRs authored by agents such as Codex are reviewed and merged more quickly than human-authored ones. However, their analysis also reveals that agentic PRs are less likely to be accepted than those of human developers, especially for complex tasks such as feature development or bug fixing. 

Complementing this investigation, another work using $567$ Claude Code-generated PRs across $157$ open-source projects, analyses how these agentic PRs differ from human ones in terms of change size and purpose~\cite{watanabe2025useagenticcodingempirical}. Their evaluation shows that agents frequently focus on non-functional improvements such as refactoring, documentation updates, and test additions. Moreover, they observed that most rejections stem from the project context rather than inherent code flows. Crucially, the work quantifies the revision effort, concluding that nearly $55\%$ of agentic PRs are merged without revision.

Nonetheless, these studies offer a limited view of agentic contributions, often focusing on a small set of PRs from a single agent or from highly popular repositories. This creates the need to study activity levels at scale across multiple agents and to investigate the ongoing evolution of agent-generated code.

\section{Methodology}
This section describes the collection, filtering, and structure of the GitHub dataset, along with the evaluation protocol. 

\subsection{Data Curation}
\label{sec:datacuration}
We target five autonomous coding agents, including 
Devin~\added{(March 2024)}~\footnote{\url{https://cognition.ai/blog/introducing-devin}}, 
Claude Code~\added{(February 2025)}~\footnote{\url{https://www.anthropic.com/news/claude-3-7-sonnet}}, 
OpenAI Codex~\added{(May 2025)}~\footnote{\url{https://openai.com/index/introducing-codex/}}, 
GitHub Copilot~\added{(May 2025)}~\footnote{\url{https://github.blog/news-insights/product-news/github-copilot-meet-the-new-coding-agent/}}, and 
Google Jules~\added{(May 2025)}~\footnote{\url{https://blog.google/innovation-and-ai/models-and-research/google-labs/jules/}}, due to their growing adoption across both academic and industrial environments, thus forming a representative coverage of current agentic coding paradigms that~\added{span} both established and more recent agents. Using the GitHub GraphQL \textit{search} API~\footnote{https://docs.github.com/en/graphql}, we scrape pull requests associated with these agents, along with other contribution metadata such as changed files, issue references, commits, comments, and reviews from each PR.~\added{To ensure coverage of all five agents,} we fix an extraction~\added{snapshot} of three months, covering June through August 2025,~\added{according to the announced release dates}.~\added{Although the time-frame is quote recent, most of these coding agents had already achieved significant usage~\cite{logicstar2025dashboard};  looking further back would have included periods when some agents did not yet exist. This window also accounts for earlier agents, such as Devin, which initially lacked PR-related capabilities~\protect\footnote{\url{https://docs.devin.ai/release-notes/overview}}\added{, and Claude Code, first released for research preview in February, before becoming generally available in May. It also includes the newest agent, Google Jules, released in late May, which quickly gained community attention through new features such as critic-augmented generation}~\protect\footnote{\url{https://jules.google/docs/changelog/2025-08-083/}}}.

To identify which pull requests correspond to which agent, we use different filtering signals according to the tell-tale signature of each agent~\added{\cite{logicstar2025dashboard, li2025riseaiteammatessoftware, watanabe2025useagenticcodingempirical}}, as illustrated in Table \ref{tab:search_signals}. When tackling a given issue, agents such as Codex and Copilot usually operate by creating a new branch prefixed with their name, followed by the issue or requirement title. Since these agents do not have an associated GitHub~\added{identity and operate as pseudo-authors}, the branch prefix serves as the~\added{primary} signal to identify related PRs, as shown in prior studies~\cite{logicstar2025dashboard, li2025riseaiteammatessoftware}. This approach focuses on the head branch, which contains all proposed changes intended for merging into the base branch. On the other hand, agents such as Google Jules and Devin operate as GitHub bots~\added{(registered as applications)},~\added{meaning} their contributions can be filtered using the author field. Lastly, instead of branch prefixing, Claude Code typically appends an authorship watermark to its contributions, which we can track in PR descriptions. 

\added{We observed that agents such as Copilot, Claude, and Codex have substantially more PRs during the selected period than the others, with Codex alone reaching approximately 1.1M PRs. As our goal is to characterize the emerging paradigm of PR-based coding agents rather than model adoption trends, we construct a representative sample for each agent covering this time window. 
According to the GitHub Innovation Graph for Q1 2025~\cite{github_developers_2025},
most GitHub developers are based in the United States, followed by the European Union, India, China, and Brazil. 
To ensure a fair comparison and minimize temporal biases, our scraping traverses the entire three-month period. We adapt the sampling granularity based on the agent's PR volume, using denser sampling for agents with higher PR activity and sparser sampling for those with fewer PRs. In practice, this corresponds to shorter time intervals (e.g., one hour for Codex) and longer intervals (e.g., five hours for Claude and Copilot), which we empirically tuned to achieve balanced coverage throughout the day. To manage days with unusually high activity, we impose a daily upper limit on sampled PRs for each agent, derived from the per-agent total sample (with respect to the least active ones), divided by the number of days. This ensures a reasonable and comparable sample size across all agents, preventing highly active ones from dominating the dataset. To maximize coverage throughout the window, the daily limit is recalculated after each day according to the remaining PRs and the number of days left, ensuring that sampling remains distributed throughout the full period rather than concentrated on a few high-activity days. Due to their lower PR activity and lack of pronounced usage concentration, we included all PRs for Jules and Devin throughout the period.}

\begin{table}[t]
\caption{Search signals~\added{for distinguishing agent- and human-generated PRs}.}
\label{tab:search_signals_human}
\Description{Search signals for distinguishing agent- and human-generated PRs}
\centering
\begin{tabular}{p{0.13\textwidth} >{\centering\arraybackslash}p{0.30\textwidth}}
\toprule
\textbf{Coding Agent} & \textbf{Search Signal} \\ \midrule
\textbf{OpenAI Codex} & head:codex/ \\
\textbf{GitHub Copilot} & head:copilot/ \\
\textbf{Cursor} & head:cursor/ \\
\textbf{Cosine} & head:cosine/ \\
\textbf{Tembo} & head:tembo/ \\
\textbf{OpenHands} & head:openhands \\ 
\textbf{Google Jules} & author:google-labs-jules[bot] \\
\textbf{Devin} & author:devin-ai-integration[bot] \\
\textbf{Amazon Q} & author:amazon-q-developer[bot] \\
\textbf{Jetbrains Junie} & author:jetbrains-junie[bot] \\ 
\textbf{Junie EAP} & author:junie-eap[bot] \\
\textbf{Codegen} & ``About Codegen'' \\
\textbf{OpenHands} & (``Co-authored-by: openhands'' OR ``Automatic fix generated by OpenHands'') \\
\textbf{Claude Code} & (``Co-Authored-By: Claude'' OR ``Generated with Claude Code'') \\
\bottomrule
\end{tabular}
\end{table}

\begin{table*}[t]
\caption{Dataset summary of pull requests and related contributions. The first three columns indicate the number of PRs created by each author, along with their base and head repositories (hence the large number), while the other four reflect activity associated with those PRs.}
\label{tab:dataset_overview}
\Description{Dataset summary of pull requests and related contributions. The first three columns indicate the number of PRs created by each author, along with their base and head repositories (hence the large number), while the other four reflect activity associated with those PRs.}
\centering
\begin{tabular}{p{0.13\textwidth} >{\centering\arraybackslash}p{0.08\textwidth} > {\centering\arraybackslash}p{0.12\textwidth} >{\centering\arraybackslash}p{0.11\textwidth} >{\centering\arraybackslash}p{0.11\textwidth} >{\centering\arraybackslash}p{0.09\textwidth} >
{\centering\arraybackslash}p{0.08\textwidth} >{\centering\arraybackslash}p{0.12\textwidth}}
\toprule
\textbf{PR Author} & \textbf{\#PR} & \textbf{\#Repository} & \textbf{\#Commit} & \textbf{\#Comment} & \textbf{\#Review} &\textbf{\#Issue} & \textbf{\#Changed File} \\ \midrule
\textbf{OpenAI Codex} & 20,835 & 41,669 & 27,530 & 3,693 & 1,957 & 45 & 90,822 \\
\textbf{Claude Code} & 19,148 & 38,260 & 82,755 & 22,329 & 12,728 & 4,052 & 255,275 \\
\textbf{GitHub Copilot} & 18,563 & 37,125 & 69,896 & 26,664 & 20,665 & 9,744 & 158,404 \\
\textbf{Google Jules} & 18,468 & 36,936 & 41,032 & 5,700 & 3,249 & 2,185 & 138,610 \\
\textbf{Devin} & 14,045 & 28,090 & 51,641 & 27,518 & 6,901 & 294 & 131,454 \\
\hline
\textbf{Human} & 20,910 & 41,542 & 102,037 & 18,559 & 21,401 & 1,973 & 194,861 \\
\bottomrule
\end{tabular}
\end{table*}

\begin{figure*}[t]
    \centering
    \includegraphics[width=0.95\linewidth]{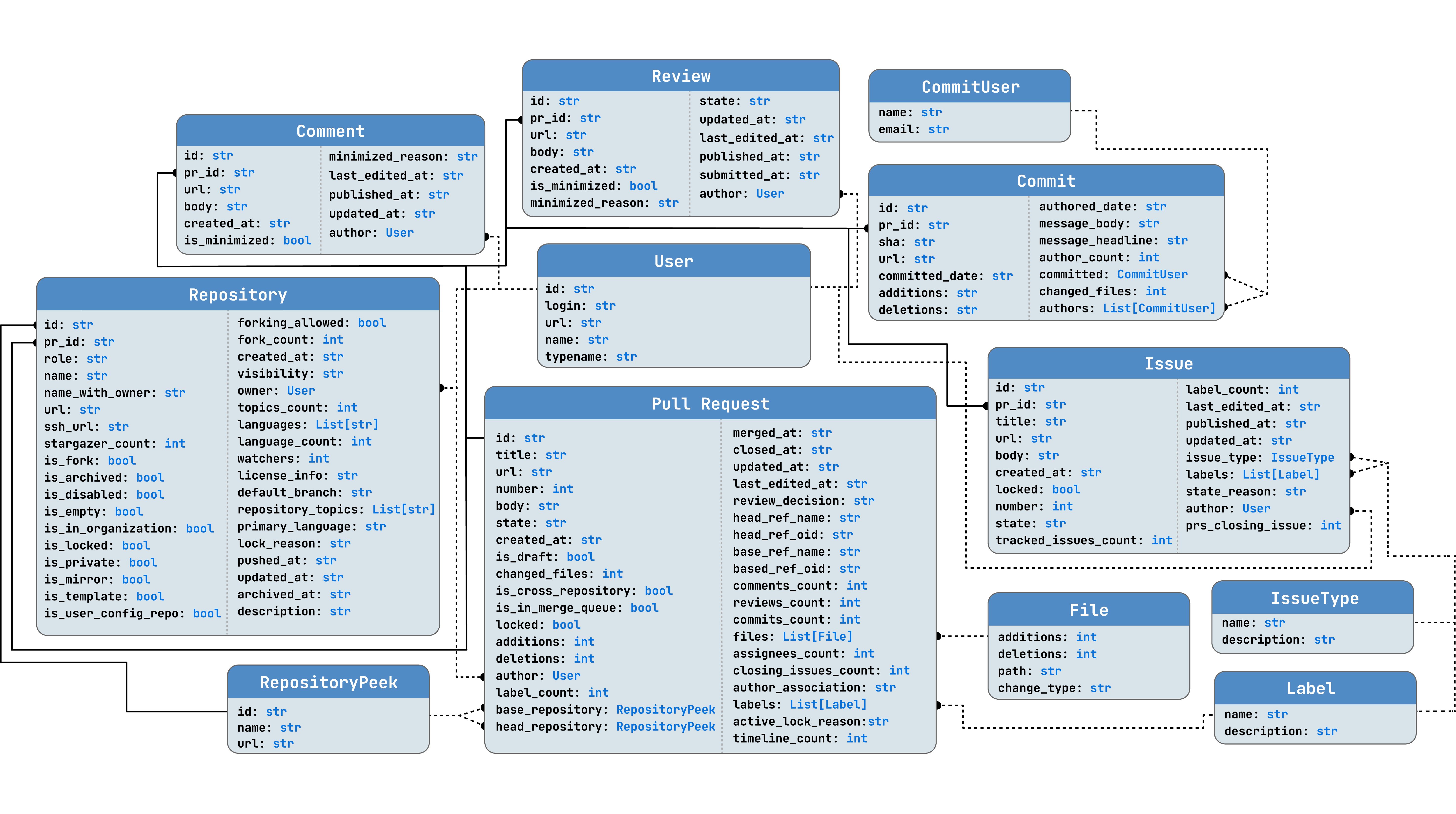}
    \caption{Overview of dataset structure. Solid lines indicate relationships between entities. Dotted lines denote nested objects.}
    \Description{Overview of dataset structure. Solid lines indicate relationships between entities. Dotted lines denote nested objects.}
    \label{fig:dataset_structure}
\end{figure*}

After~\added{collecting all agent-associated PRs} and their corresponding repositories (including base and head repositories),~\added{we performed manual sanity checks and} removed duplicate PRs\footnote{less than 0.001\% of data} based on their IDs,~\added{accounting} for repeated entries~\added{introduced by the} pagination process. In addition to the absolute counts of PR properties, we also gathered~\added{auxiliary} metadata for~\added{up to the first} 100 commits, comments, issues, reviews, and modified files associated with each PR,~\added{when available}.~\added{This limit corresponds to GitHub's maximum page size and avoids rate-limit overhead from nested pagination in the case of unusually large pull requests. Inspection of the distributions also shows that approximately 75\% of PRs (across all agents) contain between 2 and 14 entries per metadata type, supporting our threshold choice for capturing complete information in most cases while still accommodating larger PRs.} 

For comparison, we apply the same~\added{scraping procedure} to obtain human-created PRs~\added{for the same three-month window, using a one-hour time interval because of the high PR density}. To exclude agent-generated PRs, we filter based on the head branch prefix, authorship, and the presence of agent watermarks in the PR description. Because there is no clear picture of all agents that operate on GitHub,~\added{we excluded a set of known agents~\cite{logicstar2025dashboard}, as well as two additional agents, Amazon Q and JetBrains Junie, and an extra signal for OpenHands, identified through manual GitHub inspection to improve pull request attribution.} This list of agents is shown in Table \ref{tab:search_signals_human}, together with the filtering signals used for each of them. To further~\added{reduce the risk of misattributing pull requests}, we exclude PRs authored by bots or~\added{applications} (e.g., \textit{dependabot} or \textit{github-actions bot}), which are commonly used for automatic dependency management or other CI/CD~\added{tasks}. 

\subsection{Data Structure}

Table \ref{tab:dataset_overview} presents the final number of pull requests per category, together with associated activity information for each PR. The resulting dataset includes 111,969 PRs contributed by both agents and humans. Some variation in activity levels can be seen across agents, with Copilot and Claude PRs appearing more active in the collected samples. Figure \ref{fig:dataset_structure}  illustrates the overall dataset structure, which is described in more detail below.

\begin{itemize}
    \item \texttt{Pull Request}: records the content, state, and activity of a pull request, including author, repository references, timestamps, and total number of commits, reviews, comments, closed issues, labels, and files changed.
    \item \texttt{Repository}: stores a repository's ownership, visibility, status flags, popularity metrics, programming languages, topics, licensing, timestamps, and descriptive information.  
    \item \texttt{Commit}: captures a commit's identity, content, timestamps, authoring and committing information, changed files, and associated authors for a given pull request. 
    \item \texttt{Review}: lists a pull request review, including its identifier, author, content, state, timestamps, and minimization status.
    \item \texttt{Comment}: represents a pull request comment with its identifier, author, content, timestamps, publication status, and minimization details.
    \item \texttt{Issue}: stores information about an issue linked to a pull request, including its identifier, author, title, description, state, timestamps, type, labels, and other associated PRs.  
    
\end{itemize}

We also record nested information about the files changed by each~\added{pull request}. For each file, we store the number of changes in terms of additions and deletions, the type of change, and the absolute path within the repository. Similarly, for users, we capture their public-facing data such as identifier, name, login, profile URL, and user type, which~\added{were used for attribution filtering and data validation.}

\subsection{Evaluation Metrics}
\label{sec:evalmetrics}
\subsubsection{Activity Metrics}

We start with discussing some of the key high-level metrics, used to address RQ1~\added{\cite{logicstar2025dashboard, watanabe2025useagenticcodingempirical}. These metrics are computed over the collected pull requests to summarize the activity characteristics of both AI agents and human contributors.} 

\paragraph{Change Size} This metric measures the magnitude of a pull request in terms of code modifications,~\added{using the \textit{additions} and \textit{deletions} attributes. It} indicates how substantial the code modifications are within a pull request.

\begin{equation}
\text{Change Size} 
= \text{Lines Added} + \text{Lines Removed}
\label{eq:change_size}
\end{equation}

The value we pull here from the GitHub API reflects the size of the change when it was first opened. The size of the change might evolve through the PR process.

\paragraph{Merge Rate}

The merge rate indicates the proportion of submitted pull requests that are successfully merged into the target branch,~\added{using the \textit{state} attribute.}
\begin{equation}
\text{Merge rate} = \frac{\text{Number of merged PRs}}{\text{Total number of PRs}}
\label{eq:merge_rate}
\end{equation}

As noted in previous investigations, merge rates are strongly influenced by the level of human approval involved in the workflow. For example, GitHub Copilot 
can open draft PRs on its own, while Codex does so only after human approval~\cite{logicstar2025dashboard}.  

\paragraph{Merge Time}
This metric calculates the duration between the opening and merging of a pull request based~\added{on the \textit{created\_at} and \textit{merged\_at} timestamps.} It reflects how quickly a contribution is reviewed and accepted into the code base.  

\begin{equation}
\text{Merge Time} = t_{\text{merged}} - t_{\text{opened}}
\label{eq:merge_time}
\end{equation}

\paragraph{Others} We also consider other PR-related measures: the number of comments~\added{(\textit{comments\_count})} and reviews~\added{(\textit{reviews\_count})}, repository popularity~\added{(\textit{stargazer\_count})}, the fraction of changes that are additions~\added{(PR \textit{additions})}, and the variety of file types modified~\added{(\textit{path})}.

\subsubsection{Code Change Metrics}

\added{We assess code changes using the metrics defined below, applied to commits from both AI agents and human contributors, in line with RQ2.}

\paragraph{Survival Rate}

The survival rate measures the proportion of code lines that remain unchanged over a period of time.~\added{While this metric was previously applied at the repository scale~\cite{ait2022_survivalrate}, we adapt it to the commit level.} Given a seed commit, we extract all line additions in its diff. Next, we identify the nearest subsequent commit within a defined time window. Looking forward from the initial to the target commit, we capture how the newly introduced lines persist in time.  
\added{We measure survival at the syntactic level: a line is deemed surviving if its content is unchanged in the target commit, and dead otherwise.}

\begin{equation}
\text{Survival rate} = \dfrac{\text{Survived lines}}{\text{Added lines} }
\end{equation}
A higher survival rate means that the added lines of code were not modified or removed, indicating that their contribution is stable. 

\paragraph{Churn Rate}

The churn rate indicates the proportion of code lines that were added or modified over a period of time~\added{\cite{1553571}}. Similarly, starting from an initial commit, we identify changes from subsequent commits up to a target commit, focusing on modifications in the same files modified by the seed commit. This represents the number of lines that are newly added or~\added{syntactically} differ from their initial version (Churned LOC), normalized by the total number of changed lines (Total LOC).

\begin{equation}
    \text{Churn rate} = \dfrac{\text{Churned LOC}}{\text{Total LOC} }
\end{equation}
A higher churn rate reflects code has been more reworked, which may raise concerns about maintainability and stability. 

\paragraph{Deletion Rate}
The deletion rate captures the proportion of code lines removed over a period of time~\added{\cite{1553571}}. While this is similar to churn, it focuses on the number of lines deleted from the initial commit (Deleted LOC), normalized by the total number of changed lines (Total LOC).

\begin{equation}
    \text{Deletion rate} = \dfrac{\text{Deleted LOC}}{\text{Total LOC} }
\end{equation}
A higher deletion rate reflects possible refactoring or restructuring. 

\section{Results and Discussion}

In the following, we present our results, first by analyzing the activity patterns of the coding agents and then exploring how their contributions affect maintainability signals over time.

\subsection{RQ1: Agents' Contributions}

\subsubsection{Pull Request and Repositories Characteristics} We start by describing the ~\added{types} of~\added{repositories} that appear in our sample.

\begin{figure}[t]
  \centering
  \includegraphics[width=0.90\columnwidth]{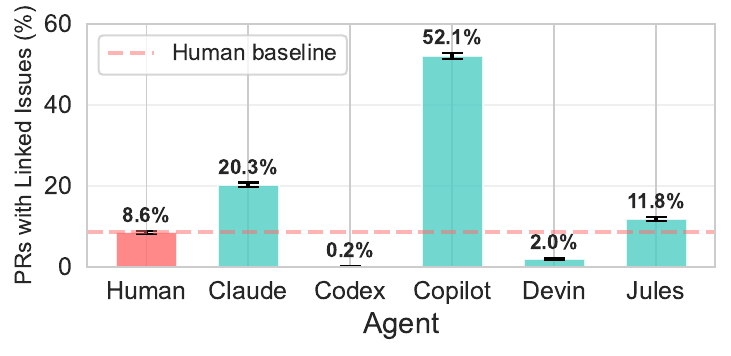}
  \caption{\added{Fraction of PRs linked to issues. Wilson 95\% CIs shown.}}
  \label{fig:issues}
  \Description{Fraction of PRs linked to issues. Wilson 95\% CIs shown.}
\end{figure}

\begin{table}[h]
\centering
\caption{Distribution of pull requests across repository stars, with most PRs originating from repositories with $0$ stars.~\added{Human} PRs tend to appear in higher-starred repositories.}
\label{tab:repo_stars}
\Description{Distribution of pull requests across repository stars, with most PRs originating from repositories with $0$ stars. Human PRs tend to appear in higher-starred repositories.}
\begin{tabular}{l r r r r r}
\toprule
\textbf{PR Author} & \multicolumn{5}{c}{\textbf{PR's~\added{Repository} Star Count}} \\
\cmidrule(l){2-6}
 & \textbf{\%=0} & \textbf{\%$\geq$10} & \textbf{\%$\geq$100} & \textbf{Median} & \textbf{Med(if$>$0)} \\
\midrule
Claude & 51.7\% & 19.7\% & 10.4\% & 0 & 5 \\
Codex & 75.3\% & 5.2\% & 2.4\% & 0 & 1 \\
Copilot & 59.6\% & 15.2\% & 9.1\% & 0 & 3 \\
Devin & 64.1\% & 23.0\% & 14.3\% & 0 & 39 \\
Jules & 75.7\% & 4.0\% & 1.4\% & 0 & 1 \\
\hline
Human & 40.5\% & 36.8\% & 25.0\% & 1 & 38 \\
\bottomrule
\end{tabular}
\end{table}

\textbf{GitHub~\added{Repository} Stars:} The number of stars~\added{of a repository} is a signal of~\added{its} popularity, community engagement, and maturity. We explore some of the details of this in Table~\ref{tab:repo_stars} for the BASE (i.e., the merged into)~\added{repositories}. We observe that \hardc{40.5\%} of~\added{the h}uman PRs in our sample are from~\added{repositories} with 0 stars. Agent PRs are more likely to be in 0-star~\added{repositories}. In the case of Codex and Jules,  about 75\% of PRs in our sample have 0 stars. While PRs are traditionally used in popular repositories for collaborating with other developers, the rise of agents is reshaping the PR usage to also be in developers' personal or small repositories.

While it is more rare that agent PRs occur in very popular 100+ star~\added{repositories}, it does occur in some cases, approaching about 15\% of an agent's PRs. The overall median stars are close to 0.~\added{Although} interestingly, when taking the median excluding 0-star~\added{repositories}, the Devin median more closely matches that of the~\added{h}uman PRs. We might speculate that because Devin is a slightly older and relatively more expensive and specialized tool, there is a larger proportion of cases where it is applied to higher-starred~\added{repositories}. In contrast, PR agents like Copilot and Codex are more directly in front of every developer with less separate cost due to their GitHub and ChatGPT integration, and thus are more likely to be used in small projects.

\textbf{Issues:} Different agents encourage different~\added{patterns}. The fraction of linked issues is shown in Figure~\ref{fig:issues},~\added{with} 95\% Wilson \cite{Wilson1927ProbableIT} confidence intervals~\added{(CIs) for statistical significance.}

Linked issues are relatively uncommon in GitHub PRs, appearing in only about 8\% of human submissions. In contrast, agents such as Copilot include linked issues in roughly half of their PRs, as assigning an issue is a central aspect of the agent’s UX, though it can also be invoked through a chat interface. Meanwhile, Codex and Devin rarely have assigned issues.

\subsubsection{Pull Request Changes} 
\label{pull_requst_changes}
Next, we examine the specific changes introduced in these PRs.

\begin{figure}[tb]
  \centering
  \includegraphics[width=0.90\columnwidth]{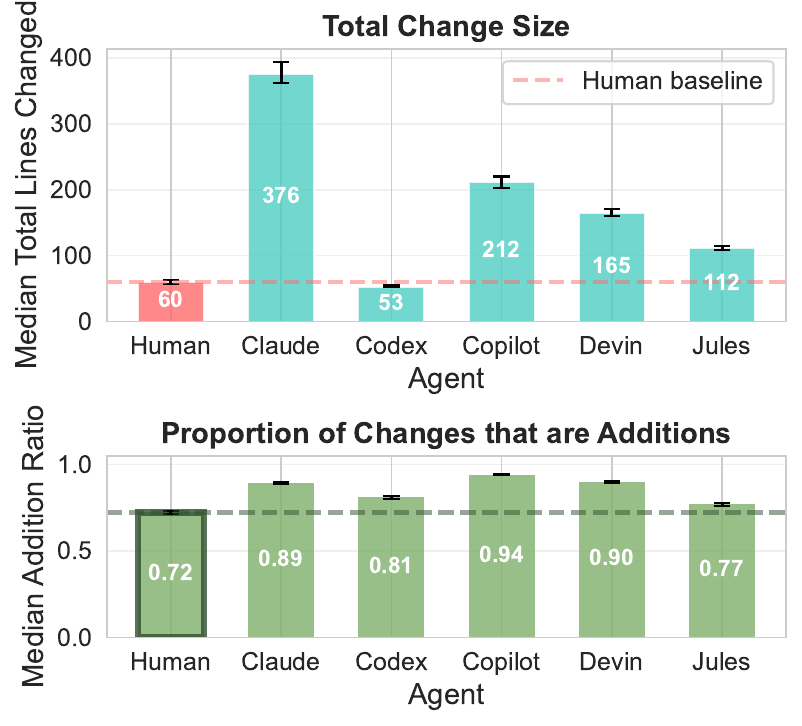}
  \caption{The plot at the top shows the median PR change size for each agent, while the bottom one \added{shows} the median fraction of changes that are additions. \added{Bootstrapped 95\% CIs shown.}}
  \label{fig:changesize}
   \Description{The plot at the top shows the median PR change size for each agent, while the bottom one shows the median fraction of changes that are additions. Bootstrapped 95\% CIs shown.}
\end{figure}

\begin{figure*}[bt]
  \centering
  \includegraphics[width=0.90\textwidth]{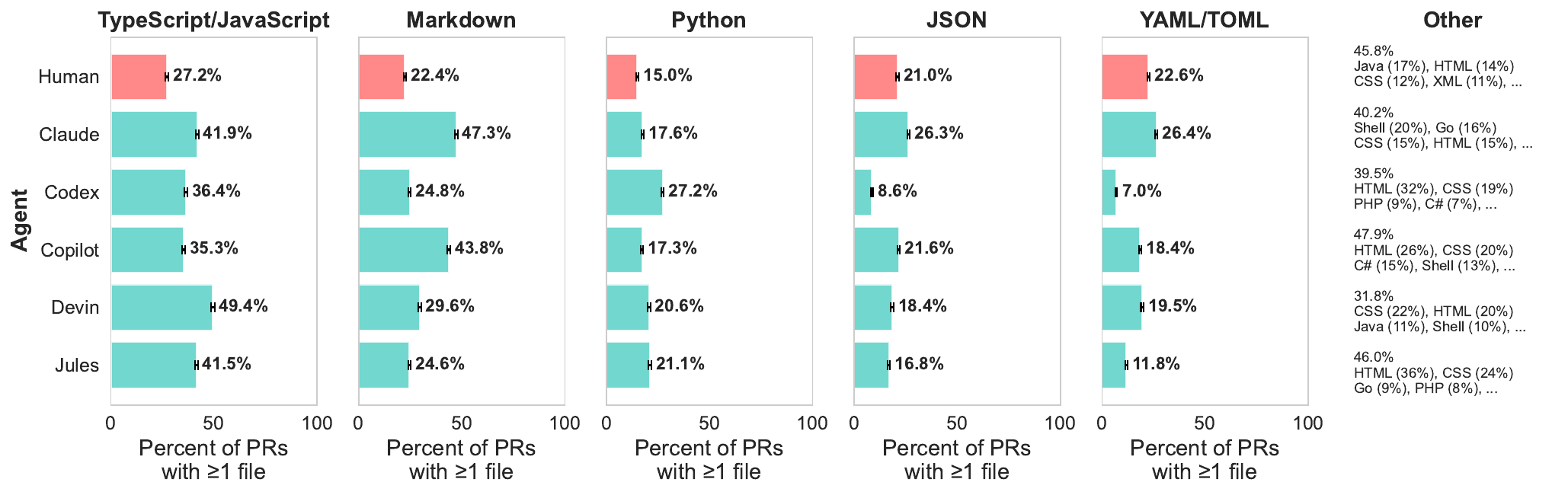}
  \caption{Percentage of pull requests containing the corresponding file type. The five most common types are shown (related types grouped, e.g., TS/JS, YAML/TOML); less common types are grouped as “Other” with right-margin annotations listing the next four most frequent types. Wilson 95\% confidence intervals are shown.}
  
\label{fig:filefacet}
\Description{Percentage of pull requests containing the corresponding file type. The five most common types are shown (related types grouped, e.g., TS/JS, YAML/TOML); less common types are grouped as “Other” with right-margin annotations listing the next four most frequent types. Wilson 95\% confidence intervals are shown.}
\end{figure*}

\textbf{Change Size:}~\added{Figure}~\ref{fig:changesize} shows summary statistics~\added{for} the number of~\added{lines} added or deleted in sampled PRs connected to~\added{each} agent. In standard diff~\added{format}, a modified line is ~\added{recorded} as both a deletion and an addition, so it counts twice toward the total lines changed.

We observe that in our sample, agent edits have progressed beyond just simple edits. The median human\added{-}opened PR was approximately~\added{60} lines. Generally agents were used in PRs with more lines changed \added{(Claude median 376 lines)}, with the exception of Codex, which was used in smaller changes. This is~\added{one of the} indicators of how Codex is often used~\added{differently} than some of the other agents.

\added{Our large sample size lends high statistical power, making differences significant\protect\footnote{A pairwise Mann-Whitney U test of change size differences estimates all $p \ll 0.0001$}. 
However, there is fairly high variance, so the actual effect size can be moderate to fairly negligible. 
A Cliff's Delta ($\delta$) \cite{Cliff1993DominanceSO} between Claude and human change sizes is about $\delta=0.37$ and about $\delta=-0.03$ between human and Codex.} For all agents, we observe higher rates of adding code compared to deleting code relative to our human sample. \added{Due to the between-PR variance, effect sizes are fairly small (e.g., human-Copilot $\delta=0.26$).}

\added{Overall, these change-size statistics suggest that agent-generated PRs tend to involve larger, addition-heavy changes, although this is not every PR. The distribution has a long tail of change lines size, so we see PRs for all authors of many sizes.}

\textbf{File Type:} In Figure~\ref{fig:filefacet}, we explore what file types are appearing in the PRs. Our scraping limits to the first~\added{100} files in the PR, and reflects touched files from comparing the latest PR HEAD to the BASE when the PR was created. 

In line with the general popularity numbers~\cite{octoverse}, we observe that JavaScript and TypeScript (TS/JS) are the most likely group of file types to appear in observed PRs.~\footnote{We opt to combine JS/TS. While these are not equivalent languages, they both reflect closely related ecosystems and usage patterns.} 
They occur in~\added{approximately} a \hardc{quarter} of~\added{h}uman PRs, but over a \hardc{third} of agent PRs. In particular, Devin has a high fraction of TS/JS PRs.

Python is the second most common programming language~\added{representing \hardc{15\%}}~\added{of} human PRs. Agent PRs are more likely to have Python code than human PRs, particularly Codex~\added{showing a} high percent of Python PRs (\hardc{2\added{7}\%}).

We also observed that~\added{a}gents frequently make changes to non-code files. Markdown file edits appear in close to half of Claude PRs, with all agents having a higher occurrence of Markdown than in human PRs. This can be a reflection of the propensities of agents to add documentation, but also might reflect that the agents are more likely to be used for "$0\to1$" type edits, where creating new~\added{repositories} or~\added{independent features} from scratch. While these are the most common file types to be included, there is still a long tail of other file types included in PRs.

\subsubsection{Pull Request Merges} 
Next we explore what happens to these PRs after being opened.
\label{pullrequestmerges}

\begin{figure*}[tb]
    \centering
    \includegraphics[width=0.90\textwidth]{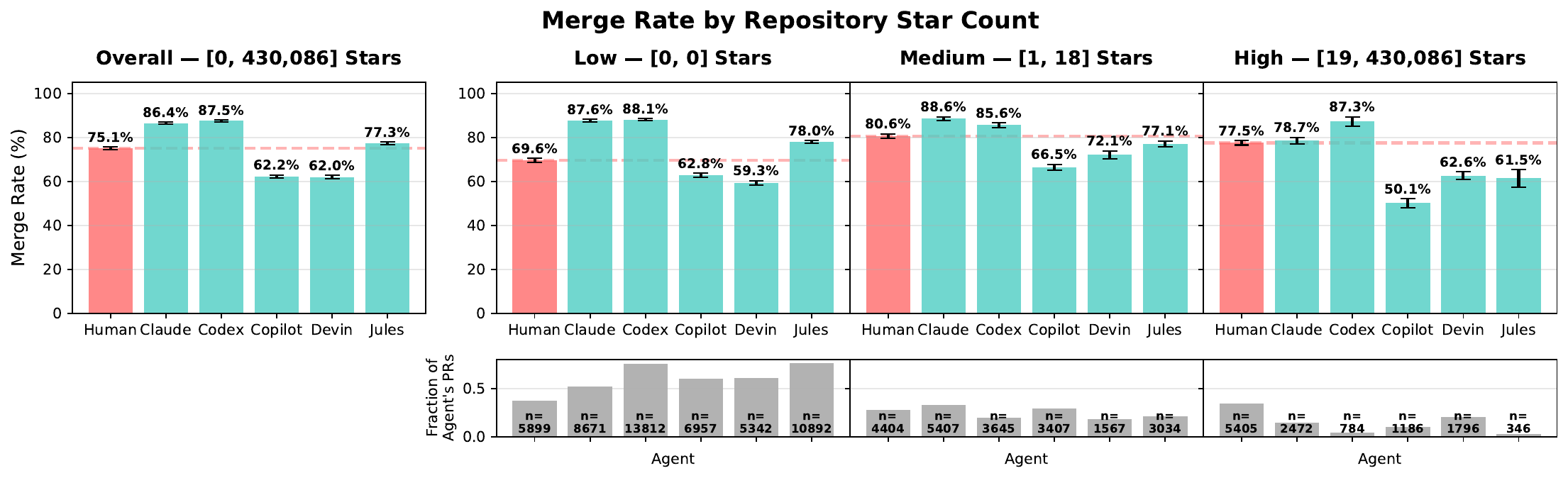}
    \caption{Percentage of pull requests merged per agent, grouped by repository star count (Wilson 95\% CIs shown). Lower panel shows each agent's PR fraction within star categories (p33 and p66 of human data). Some agents show an “inverted-u” trend across star counts.}
    \label{fig:mergeratestrat}
    \Description{Percentage of pull requests merged per agent, grouped by repository star count (Wilson 95\% CIs shown). Lower panel shows each agent's PR fraction within star categories (p33 and p66 of human data). Some agents show an “inverted-u” trend across star counts.}
\end{figure*}

\begin{figure*}[tb]
    \centering
    \includegraphics[width=0.90\textwidth]{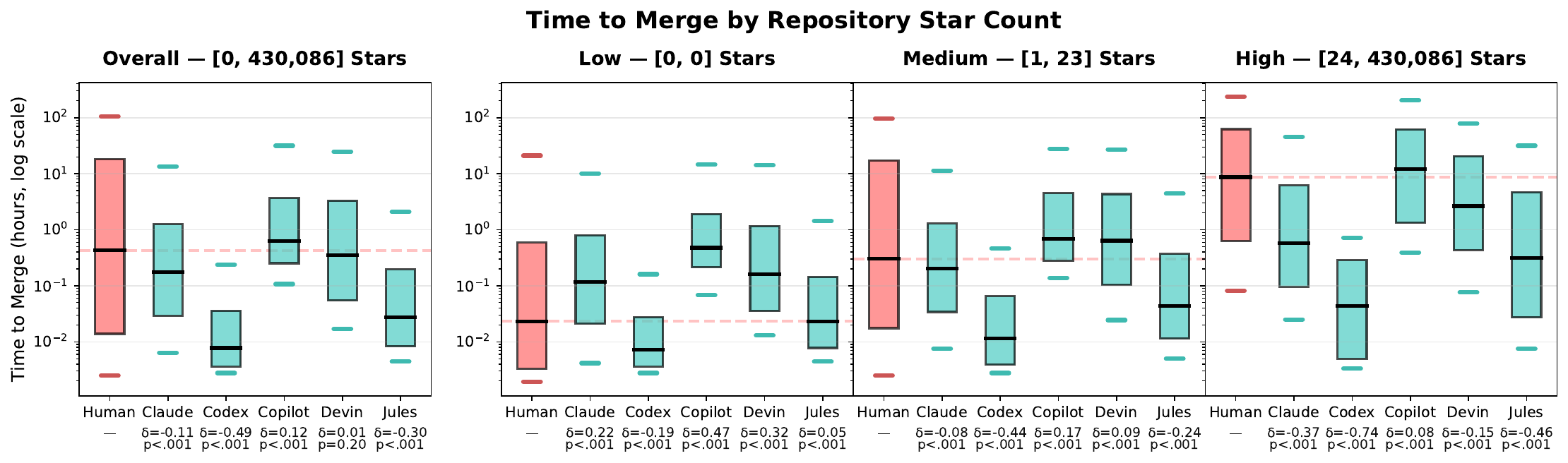}
    \caption{Distribution of merge times across agents (log scale) as boxplots with p10–p90 whiskers. Cliff's $\delta$ and Mann-Whitney U test p-values comparing to humans are labeled. Merge times generally increase with repository star count; Codex and Jules PRs merge faster, while shorter times may reflect prior human interaction. Star boundaries are based on human PR tertiles of merged PRs.}
    \label{fig:mergetimestrat}
    \Description{Distribution of merge times across agents (log scale) as boxplots with p10–p90 whiskers. Cliff's $\delta$ and Mann-Whitney U test p-values comparing to humans are labeled. Merge times generally increase with repository star count; Codex and Jules PRs merge faster, while shorter times may reflect prior human interaction. Star boundaries are based on human PR tertiles of merged PRs.}
\end{figure*}

\textbf{Merge Rate:} In Figure~\ref{fig:mergeratestrat},~\added{we compare merge rates across agents, shown both for the full dataset and stratified by repository popularity (right), grouped into three star-count tertiles. Each PR can be in one of three states: open, merged, or closed. We show the percentage of all PRs that are merged out of all PRs (as PRs that are left open can represent cases where agents did not resolve an issue, but caused developer effort). }

Across all the data, we observe that agents such as Claude and Codex are merged at a higher rate than human PRs in our sample, while Copilot and Devin are merged at a lower rate (\added{we note significantly separated CIs in Figure~\ref{fig:mergeratestrat}}). When looking at the star count, we observe different patterns depending on the PR author. Human, Copilot, and Devin PRs follow a slight ``inverted-u'' pattern; the 1st tertile and 3rd tertile~\added{have} lower merge rates than the 2nd. The merge rate for Codex is roughly similar across star count\added{s}. Jules merge rate is fairly flat but falls off at higher star counts (notably, Jules is also the agent where our sampling finds the fewest \added{PRs from high-starred repositories}).

\begin{figure}[tb]
  \centering
  \includegraphics[width=\columnwidth]{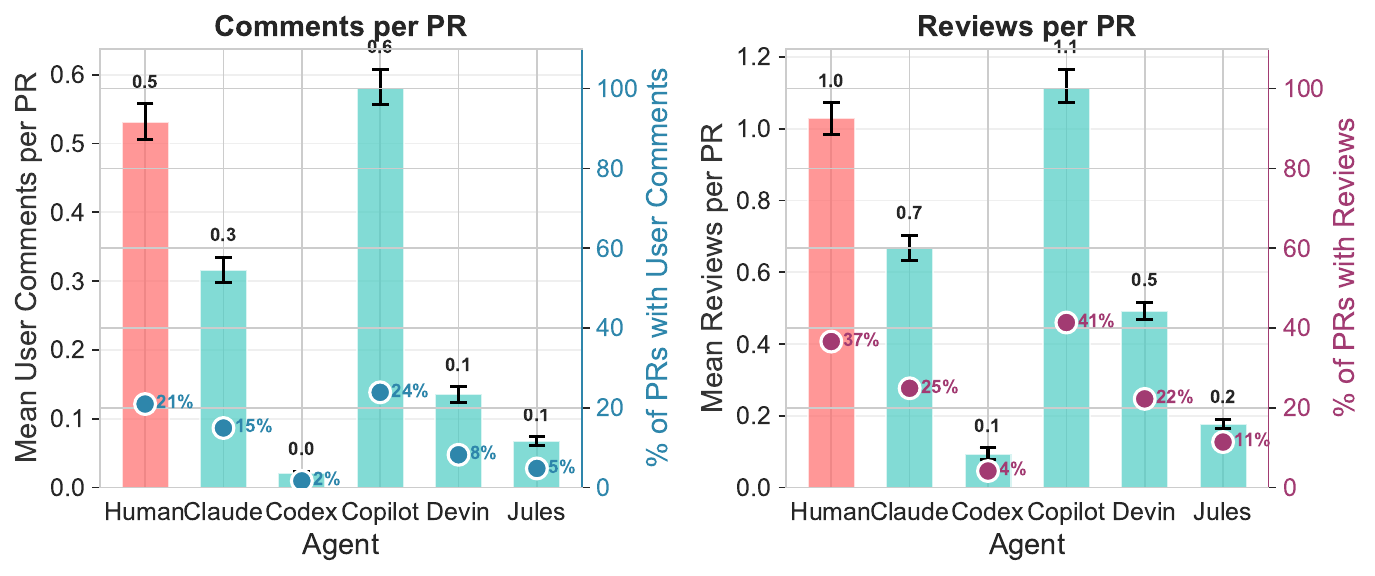}
  \caption{Developer interactions overview with pull requests, showing average counts of user comments (i.e., non-bot comments), reviews (main bar, left axis), and the proportion of PRs with at least 1 comment or review (dots, right axis).  
  \added{Bootstrapped 95\% CIs shown.}
  }
  \label{fig:mainengage}
  \Description{Developer interactions overview with pull requests, showing average counts of user comments (i.e., non-bot comments), reviews (main bar, left axis), and the proportion of PRs with at least 1 comment or review (dots, right axis). Bootstrapped 95\% CIs shown.}
\end{figure}

\textbf{Merge Time:} In Figure~\ref{fig:mergetimestrat} we show the distribution of times between~\added{PR opening and merge events}. A typical PR in our sample is merged fairly quickly. The median human PR is merged in \HumanMedianTimeToMergeHours hours, and a large fraction (\HumanPctMergedUnderTenMin) of PRs are merged in under 10 minutes. These ranges are somewhat typical for other agents except for Codex and Jules. The median Codex PR in our sample was merged in \CodexMedianTimeToMergeMinutes minutes.

If we consider merge times grouped into three categories by repository star count (right side of Figure~\ref{fig:mergeratestrat}), we observe an increasing trend. In the case of humans, the median PR is merged in only \hardc{a few minutes}. This increases to a median of about \hardc{10 hours} in higher starred~\added{repositories}.~\added{Although the} merge time for some agents' PRs (\added{such as} Copilot and Devin) also shifts up with higher starred~\added{repositories}, the p75 of merge times for Codex remains under 1 hour even for higher starred~\added{repositories}.

\added{Consistent with the change size findings, this offers an additional perspective} on how Codex (and to some extent Jules) is being used differently than other agents. 
We~\added{speculate} that the faster merging might arise from~\added{a}gent PRs
performing much simpler (or rote) work, or that the~\added{a}gent code is reviewed
by humans, off-line, before PR submission, and using PRs only as the AI system's provided mechanism for integrating code into the codebase. The distribution of Copilot PR merge time in high\added{-starred}~\added{repositories} is surprisingly similar to~\added{h}uman PRs (thus\added{,} they might be receiving similar amounts of review effort), but there is an approximate 27 points gap in merge rates.

\textbf{High-level Interaction Metrics}
In Figure~\ref{fig:mainengage}, we look at some of the measures of how developers~\added{interact} with PRs. Some agents (such as Devin) will always comment on their own PRs, while others will not. Thus\added{,} we focus on comments labeled by GitHub as being from a user rather~\added{than} a bot.

Comments are fairly rare. Only \hardc{21\%} percent of~\added{h}uman PRs in our sample receive comments. We observe that agent PRs typically receive fewer comments (with codex PRs rarely commented on). We also note that human PRs receive
fewer reviews than~\added{a}gent PRs (although Co~\added{p}ilot's PRs get marginally more reviews). In a traditional framing of PRs, this would perhaps hint that the PRs and associated changes are simpler and do not require much review. However, agent systems can alter this paradigm. For example, Codex and Jules review might have been more likely to occur off-platform in a chat interface, while Copilot is more closely tied with reviews as the mechanism to invoke the system for edits. Additionally, systems~\added{such as} Copilot include~\added{an} UI for using~\added{the system} itself as a reviewer.

\subsection{RQ2: Code Change over Time}
\label{sec:codechangeovertime}
We examine code evolution at three post-commit intervals: three days, one week, and three weeks. We restrict our analysis to pull requests merged into the code base, as our objective is to understand how successfully-integrated code~\added{changes} over time.~\added{These time intervals were chosen to capture short-range changes following code integration, without extending the analysis into substantially longer horizons (e.g., several months or years), which would require additional longitudinal data. While code may continue to change over longer time spans, our study focuses on this early post-merge window as a snapshot of current agent-driven ~\mbox{development practices.}} 

\begin{figure}[tb]
  \centering
  \includegraphics[width=\columnwidth]{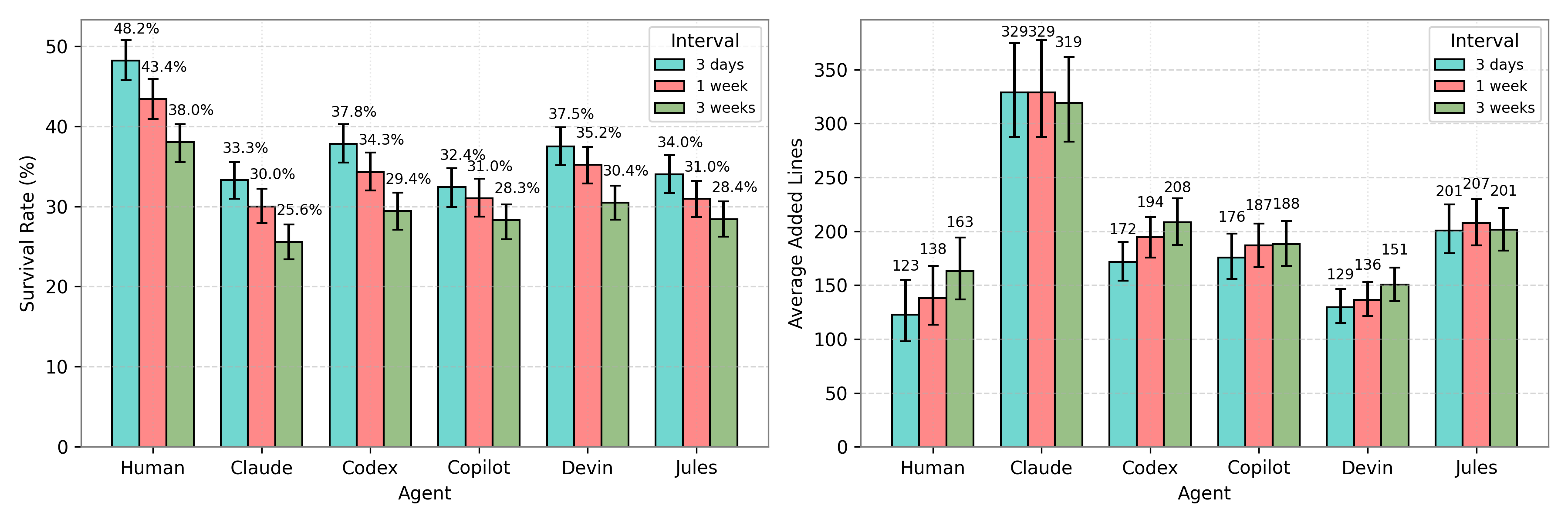}
  \caption{Left plot denotes the fraction of commits with complete line survival across agents measured at 3-day, 1-week, and 3-week intervals. The right plot shows the average net additions per agent across the same time intervals.~\added{Bootstrapped 95\% CIs shown.}}
  \label{fig:surv_rate}
  \Description{Left plot denotes the fraction of commits with complete line survival across agents measured at 3-day, 1-week, and 3-week intervals. The right plot shows the average net additions per agent across the same time intervals. Bootstrapped 95\% CIs shown.}
\end{figure}

\begin{figure}[tb]
  \centering
  \includegraphics[width=\columnwidth]{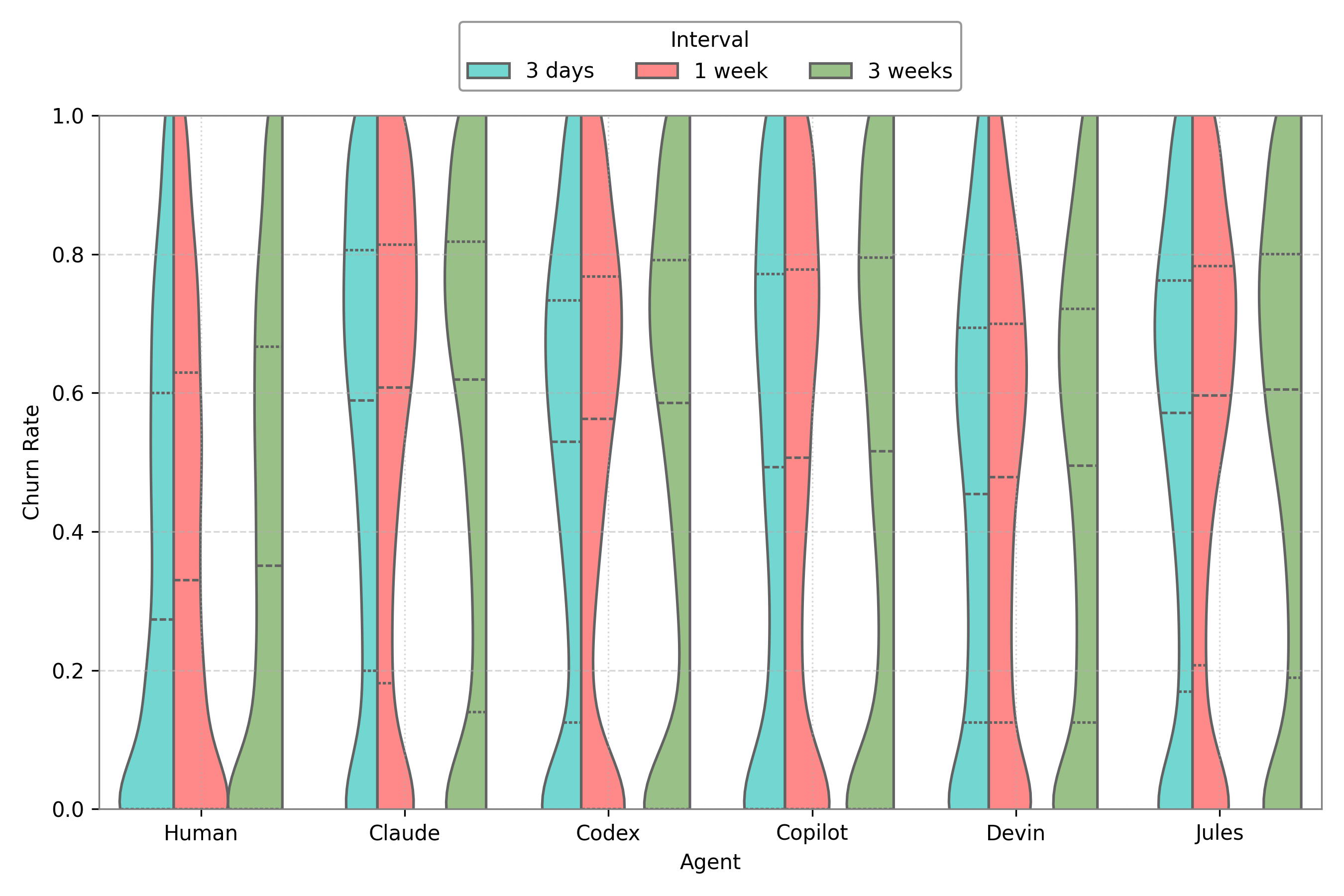}
  \caption{Per-agent distribution of commit churn rates measured over 3-day, 1-week, and 3-week intervals.}
  \label{fig:chrn_rate}
  \Description{Per-agent distribution of commit churn rates measured over 3-day, 1-week, and 3-week intervals.}
\end{figure}

~\added{Furthermore, accurately identifying the author of commits can be challenging for some agents.} For example, Claude Code allows its watermarks to be disabled, while OpenAI Codex often records commits under developers' accounts~\added{\cite{li2025riseaiteammatessoftware}}.

To~\added{minimize uncertainty and ensure comparability, we consider the commit reflecting the agent’s initial implementation of the assigned issue, typically the first commit in the PR. Since we know the PR authorship to be agent-generated from our filtering of observable signals, this commit represents the agent's first contribution to the PR prior to review, reducing the risk of confounded authorship.} 

However, we note that for GitHub Copilot, the first commit in a PR is commonly a placeholder used to initialize an issue plan without introducing any file changes. To address this, we instead select the second commit from each Copilot PR, which reflects the actual implementation of the requirement. On top of that, we exclude any commits with no file changes across all agents, if any. 

Finally, we sample $1560$ commits for each agent and for humans. To reduce bias from repositories with zero stars, we perform stratified random sampling based on repository popularity. Repositories with zero stars form a distinct category, while those with at least one star are sampled according to quantile-based star distributions. 

\paragraph{Code Survival}

In Figure~\ref{fig:surv_rate}, the left plot shows the percentage of commits where all their initial additions have survived, across all agents. For all three intervals, we can see that the fraction of commits where all lines survived is consistently higher for humans than for any agent, with nearly half of their additions surviving to three days. ~\added{Mann-Whitney U tests show these differences are statistically significant ($p < 0.001$), and Cliff's $\delta$ values (-0.05 to -0.14) suggest the effect sizes are small but consistently negative, indicating that human-generated changes might persist slightly longer}. In contrast, agent-generated commits are more likely to have their lines reworked, suggesting a lower stability of their contributions. Although the overall decline in survival over time for both humans and agents reflects the natural attrition of code, the discrepancy between them could be due to how agents are used in practice. The drop in survival for agent-generated commits may stem from their changes being more localized and repetitive. It is worth noting that commits from agents tend to produce a larger net code increase over the observation interval, as shown in the right plot of Figure~\ref{fig:surv_rate}. This may result from agents generating larger code snippets per commit. Alternatively, human commits likely involve more deliberate and distributed design decisions, which could explain their higher commit survival.

\begin{figure}[tb]
  \centering
  \includegraphics[width=\columnwidth]{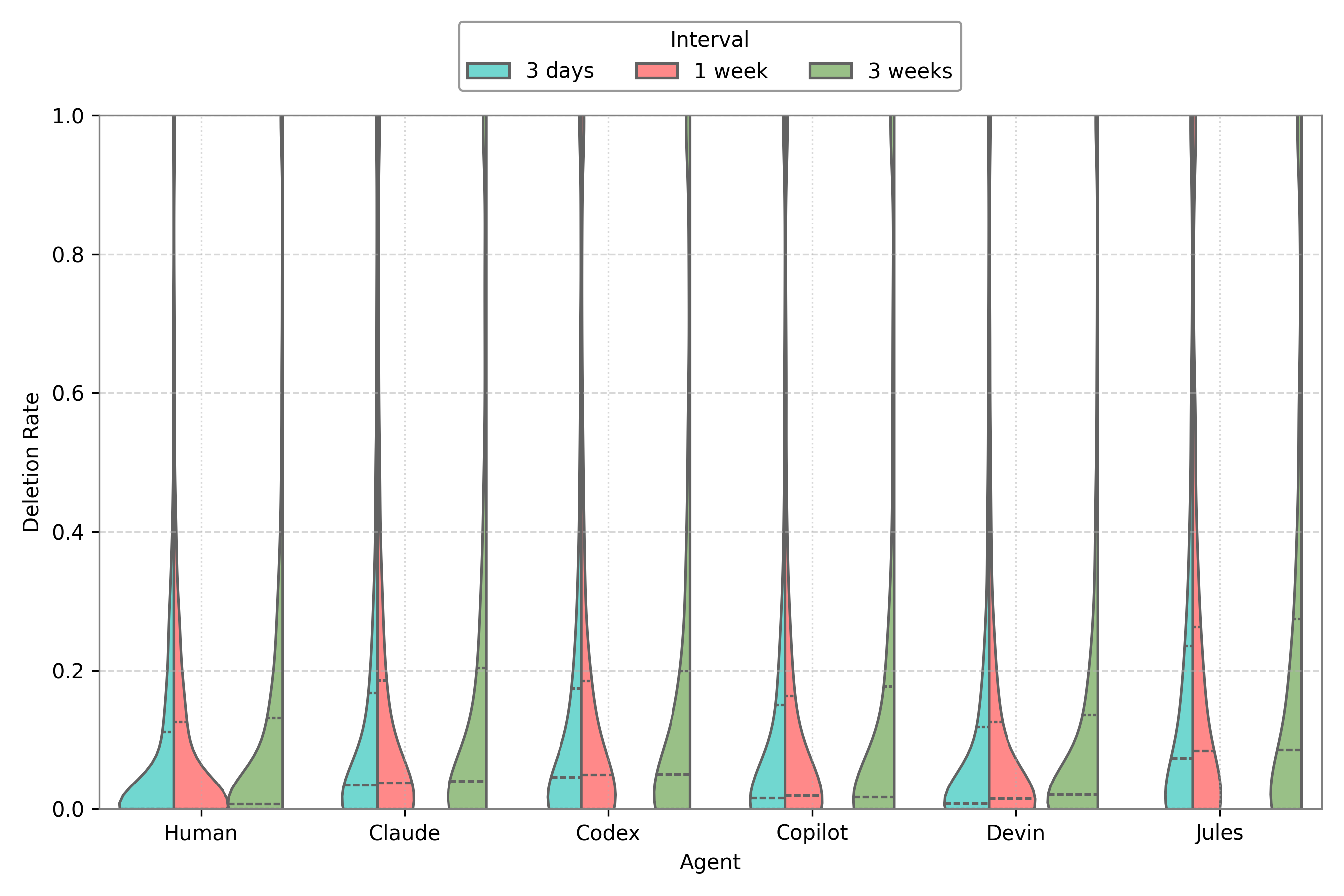}
  \caption{Per-agent distribution of commit deletion rates measured over 3-day, 1-week, and 3-week intervals.}
  \label{fig:delete_rate}
  \Description{Per-agent distribution of commit deletion rates measured over 3-day, 1-week, and 3-week intervals.}
\end{figure}

\paragraph{Code Churn}

In Figure~\ref{fig:chrn_rate}, we illustrate the distribution of commit churn rates for each agent across the three observation intervals. For humans, the distribution is concentrated mainly between $0$ and $0.4$, indicating that most commits introduce relatively few changes to the total modified files. This suggests that overall human contributions are more stable than agent ones, with less volatility in code modifications. On the other hand, agent commits display a higher median churn compared to humans across all time periods, indicating that their changes are more frequent and larger in proportion to the total changed lines. This tendency is more evident in Claude, which shows pronounced concentration of churn values between $0.8$ and $1$. This behavior could be associated with the tendency of Claude Code to modify larger portions of the code base in each commit, as we have seen in Section ~\ref{pull_requst_changes}, which could potentially increase the occurrence of later revisions. This high number of changes could stem from formatting, documentation (such as more markdown files as seen in Figure~\ref{fig:filefacet}), or stylistic adjustments, which are common in automatically generated code. 

Furthermore, the distributions for agents themselves remain largely consistent across the three time intervals, with only a slight upward shift in median churn. This suggests that most rework occurs within the first few days after a commit, and that code tends to stabilize relatively quickly after. Compared to the subsequent decline in commit survival over longer intervals, the stable churn indicates that individual commits generally involve incremental changes, while the later survival drop reflects the cumulative effect of other commits on previously added lines rather than large-scale rework in any single commit. 

Additionally, Figure~\ref{fig:delete_rate} shows the distribution of line deletion rates per commit for each agent. Although deletion rates are low, agent commits consistently exhibit slightly higher deletion ratios than human ones, suggesting that auto-generated code is more frequently replaced or refined.~\added{Similar to the survival rate, statistical tests show a significant difference between human and agent commits in terms of churn and deletion rates across all intervals, with effect sizes remaining small ($\delta \approx$ 0.04–0.29).} Taken together, these findings indicate that while human contributions might be more stable, agent-generated commits are more susceptible to early changes and deletions. 
\added{Note that this represents only one perspective on the evolution of agent-generated code. Future efforts should examine this phenomenon in greater depth; for example, by analyzing the drivers of code churn (bug fixes, feature additions, requirements changes, or flawed abstractions), and by more clearly distinguishing the types of changes made in real-world PRs, such as code, comments, and documentation. Such analyses would further deepen our understanding.}

\section{Future Work}
 
The rapid evolution of agentic development offers numerous directions for future research. Beyond analyzing agentic activity in practice, our dataset provides a foundation for benchmarking the quality, stability, and collaboration patterns of agent-generated code. Researchers can use it to fine-tune or evaluate models on tasks such as code generation and repair, using real-world outcomes, such as post-merge churn or acceptance rates, as objective signals. The dataset can also be extended to span longer periods, include new agents, and integrate richer metadata to enable studies of behavioral changes over time and across diverse projects. Future work might also incorporate CI/CD metrics (e.g., build and test results), PR timelines (e.g., review iterations), and contributor information.
The long-term effects of agent-generated code could also be studied through analysis of technical debt accumulation and its relationship to agent versions, project characteristics, and review practices.

\section{Threats to Validity}
\label{sec:threatstovalidity}
\paragraph{System Version Changes Opaque}
This is a rapidly evolving area. Even as we scraped our data, 
these agent systems were actively being tweaked and modified. Our findings partially accommodate this by focusing on aggregated behavioral patterns across all pull requests, capturing the emergent, longitudinal behavior of agentic development rather than individual version-specific effects. 

\paragraph{Agent Contribution Transparency}
Agents such as OpenAI Codex and Claude Code can make untraceable contributions via their CLI interface, and any authorship watermarks that might signal these contributions can be disabled. Particularly for Codex, activity is usually recorded under human accounts, making it difficult to fully track more granular contributions. To account for this, our analysis focuses only on pull requests that can be reliably attributed to agents and, therefore, reflects only the activity that is observable rather than the absolute range of agent contributions. Furthermore, for the~\added{code change} analysis, we specifically select~\added{the initial} commits from merged PRs created by agents,~\added{to minimize the risk of confounded authorship, while recognizing that precise attribution remains an open challenge, especially as agent-based workflows on GitHub are still emerging.} 

\paragraph{Human Attribution Uncertainty}
Since the full set of active coding agents on GitHub is unknown, filtering for purely human-authored data may still include undetected agentic contributions. To reduce this threat,~\added{we identify PRs using signals consistent with prior work, supplemented with additional agents identified through manual inspection to improve coverage of the agentic landscape at the time of the study. We also exclude any PRs created by bots or applications that are not real users, and we apply the same commit authorship strategy used for agents. Together, these measures collectively remove most automated activities.}

\paragraph{Repository Popularity Imbalance}

\added{The collected PRs show a skewed distribution of repository popularity, particularly across agents, with most PRs originating from low-star repositories. This pattern reflects the distribution seen in the wild and aligns with prior work~\cite{logicstar2025dashboard}. In the activity analysis, we report results by repository popularity levels to show how activity patterns vary in practice. For the code change analysis, we mitigate potential bias from repository popularity by using stratified sampling across different popularity levels.} 

\paragraph{Data Collection Coverage}
For agents with a large number of PRs, the GitHub API may return contributions concentrated within specific time frames, which may lead to the dataset underrepresenting the global distribution of agentic contributions. To mitigate this, we uniformly sample PRs across days and time intervals within each day, proportionally to the total number of PRs for each agent. 

\section{Conclusion}

AI coding agents, although a recent phenomenon, are evolving rapidly and already shaping software development. Public PRs from these agents offer a valuable lens on this shift, despite UX and usage differences that make comparability between agents and humans challenging. Our longitudinal study of PRs suggests that agent-generated code is significantly less likely to be retained and more prone to churn. We also observe a dramatic rise in agent generated PRs targeting zero-star repositories. Although such PRs behave differently from the PRs in more popular repositories, they cannot be ignored, as they currently represent the majority of agent activity. Whether these agents will see wider adoption in higher-starred repositories remains uncertain.
The reasons for this pattern require further investigation, but the findings may have important implications for assessing the overall costs and benefits of automated agents. We hope that our study adds to the initial understanding of this important emerging trend.

\newpage
\printbibliography

\end{document}